\documentclass[12pt]{iopart}
\usepackage{iopams}  
\usepackage{graphicx}  

\begin{document}

\title[$\phi$ Meson Production in In-In Collision]{$\phi$ Meson Production
  in In-In Collisions and the $\phi$ Puzzle}

\author{M Floris for the NA60 Collaboration\footnote{The full list of
    authors can be found at the end of this volume.}}

\address{Universit\`a degli Studi di Cagliari and INFN Sezione di Cagliari,
strada prov.le per Sestu km 0.700, 09042, Monserrato (Cagliari)}
\ead{michele.floris@ca.infn.it}
\begin{abstract}
  The NA60 experiment measured dimuon production in In-In collisions
  at 158 AGeV. This paper presents a high statistics measurement of
  $\phi\to\mu\mu$ with the specific objective to provide insight on
  the $\phi$ puzzle, i.e. the difference in the inverse $T$ slopes and
  absolute yields measured by NA49 and NA50 in the kaon and lepton
  channel, respectively.  Transverse momentum distributions were
  studied as a function of centrality. The slope parameter $T$ shows a
  rapid increase with centrality, followed by a saturation. Variations
  of $T$ with the fit range of the order of 15~MeV were observerd,
  possibly as a consequence of radial flow.  The $\phi$ meson yield
  normalized to the number of participants increases with centrality
  and is consistently higher than the yield measured by the NA49
  experiment at any centrality.
\end{abstract}


One of the most striking predictions of QCD is the occurrence of a
phase transition from ha\-dro\-nic matter to a deconfined plasma of
quarks and gluons, when sufficiently high energy densities are
reached.  Strangeness enhancement was proposed long ago as a signature
of such phase transition~\cite{Rafelski:1982pu}. Experimentally, a
rather interesting observable is the $\phi$ meson, due to its
$s\bar{s}$ valence quark content.  Moreover, it has been argued that
close to the phase boundary, the spectral function of the $\phi$, and
thus its mass, width and branching ratios, could be
modified~\cite{Shuryak:1999zh}.  The yield and transverse momentum
spectra of the $\phi$~meson were studied extensively at the SPS in
Pb-Pb collisions at $158~\mathrm{A\,GeV}$.  The NA49 experiment
measured the $\phi \to KK$ channel~\cite{Afanasev:2000uu} (with
statistics limited to $p_T \lesssim 1.6~\mathrm{GeV}$), while the NA50
experiment measured the $\phi \to \mu \mu$
channel~\cite{Alessandro:2003gy} (with acceptance limited to $p_T >
1.1~\mathrm{GeV}$).  Both experiments observed an enhancement of the
$\phi$ yield with the size of the collision system.  However, the
absolute yields measured in the common kinematic window (transverse
momentum, rapidity and Collins-Soper angle) disagree by a factor of
about two~\cite{Jouan}.  It was suggested that kaon absorption or
rescattering prevents the reconstruction of most in-matter $\phi\to
K\bar K$ decays, in particular at low transverse momentum, while the
$\phi$ mesons decaying in the lepton channel would not be
affected~\cite{Shuryak:1999zh,Johnson:1999fv,Pal:2002aw}. The effect
predicted by these models, however, is smaller than the observed
difference.  NA49 and NA50 disagree also on the inverse $T$ slopes
determined from the transverse momentum spectra. The NA49 inverse
slopes are consistently larger and with a stronger centrality
dependence than those measured by NA50.  This could be partly due to
radial flow which tends to make the $T$ slopes extracted from a low
$p_T$ region higher.  A further increase of the NA49 slopes might be
due to the depletion of the $\phi\to KK$ at low transverse momentum.
The discrepancy in absolute yields and slopes became known as the
{$\phi$ puzzle}~\cite{Shuryak:1999zh,Rohrich:2001qi}.
Production of $\phi$ in central Pb-Pb collisions was recently studied
also by the CERES collaboration~\cite{Adamova:2005jr} both in the kaon
and in the dielectron channel. The measured inverse slope and yield
are compatible with the NA49 results in both channels, although the
$ee$ measurement is affected by large uncertainties.

This paper presents a high statistics measurement of $\phi\to\mu\mu$
in In-In collisions at 158~${\mathrm{A\,GeV}}$, with the specific
objective to provide insight into the $\phi$ puzzle.  The sample
consists of 360000 signal pairs, with $\sim$70~000 events lying in the
$\phi$~peak. The si\-gnal/back\-ground ratio integrated over
centrality is $\sim1/3$ below the $\phi$~peak.  The data were divided
in 5 centrality bins, selected on the basis of the number of
participants, with $\left< N_{part} \right> = 15, 39, 75, 132, 183$.
The number of participants was estimated from the charged tracks
multiplicity measured in the vertex detector, matching the knee at
high multiplicities to a Glauber calculation and assuming a linear
relation $N_{part} \propto N_{ch}$. For more details on the analysis
procedure and on backgrounds subtraction see
Ref.~\cite{Floris:SQM,Shahoian:2005ys}.


\begin{figure}[tbp]
  \centering
  \includegraphics[width=0.43\textwidth]{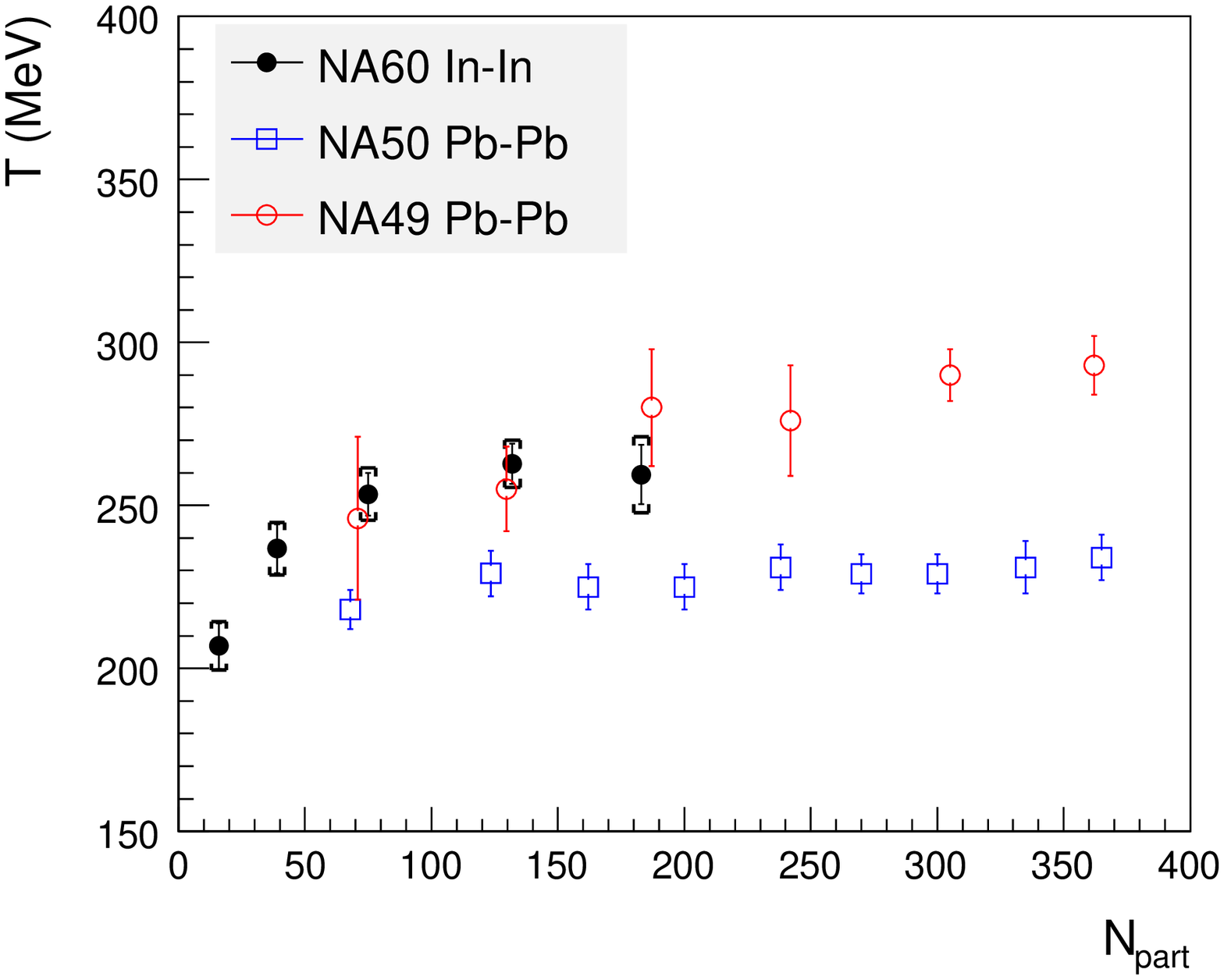}
  \hfill
  \includegraphics[width=0.43\textwidth]{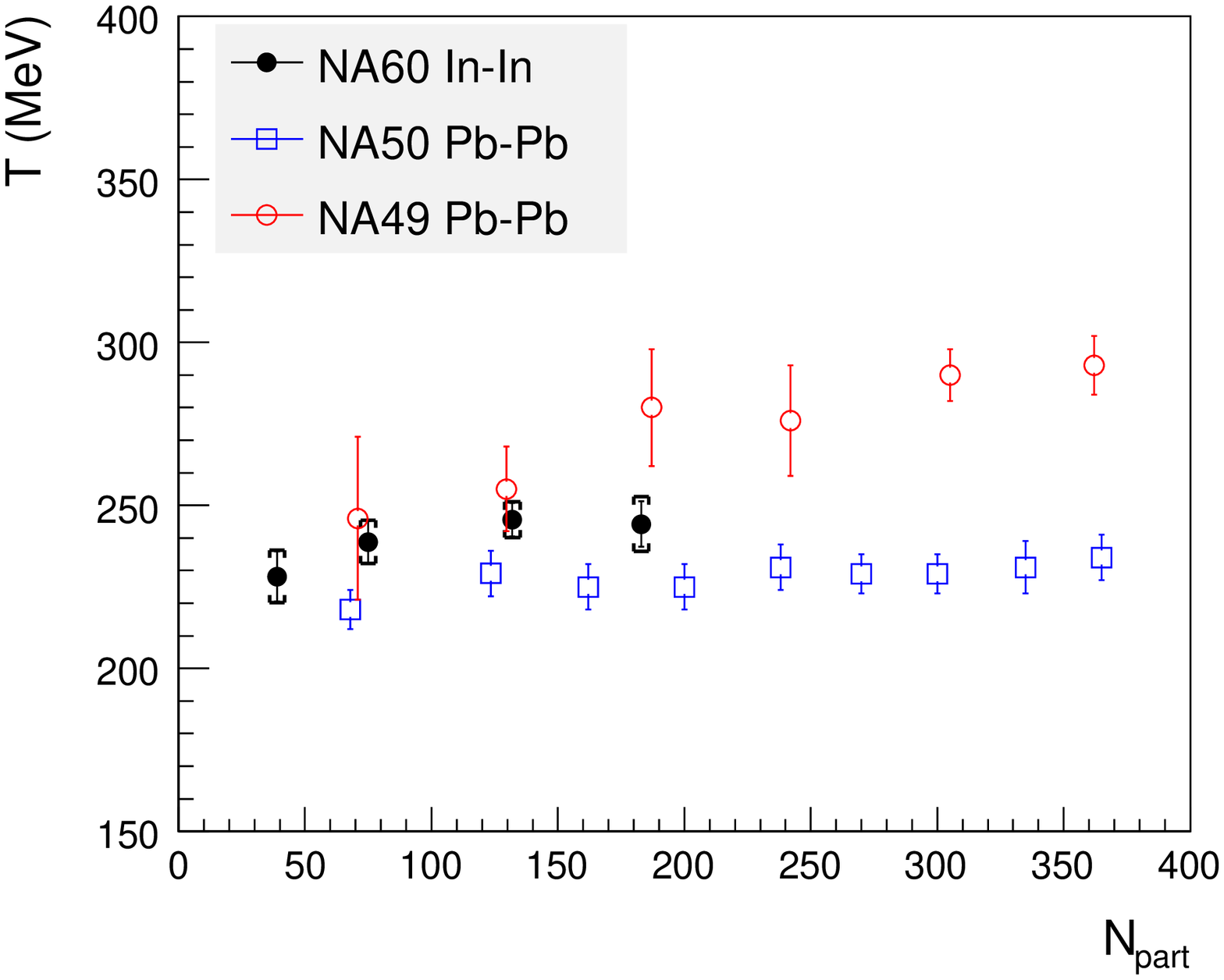}
  
  \caption{$T$ slope parameter as a function $N_{part}$ compared to
    the results of NA49 and NA50. Left: fit performed in the NA49
    range ($0 < p_\mathrm{T} < 1.6~\mathrm{GeV}$).  Right: Fit
    performed in the NA50 range ($1.1 < p_\mathrm{T} <
    2.6~\mathrm{GeV}$; due to low statistics, the 2 most peripheral
    points are integrated). Square brackets: total error (statistical
    and systematic).}
  \label{fig:tslope}
\end{figure}

In order to study the $p_{T}$ distributions, events in a small mass
window centered at the $\phi$ pole were selected ($0.98~\mathrm{GeV} <
M_{\mu\mu} < 1.06~\mathrm{GeV}$).  The spectrum of the continuum below
the $\phi$ was estimated and subtracted selecting events in two side
mass windows ($0.88~\mathrm{GeV} < M_{\mu\mu} < 0.92~\mathrm{GeV}$ and
$1.12~\mathrm{GeV} < M_{\mu\mu} < 1.16~\mathrm{GeV}$).  After
subtraction of the combinatorial background, physical continuum and
fake tracks, the kinematic distributions were corrected for acceptance
times reconstruction efficiency, estimated with an overlay Monte Carlo
simulation (a Monte Carlo dimuon is reconstructed on top of a real
event).  In principle this correction should be evaluated as a
3-dimensional matrix, as a function of $p_{T}$, $y$ and
$\cos\,\theta_{CS}$. This approach also requires fiducial cuts to
remove the phase space corners, with a corresponding loss in
statistics.  In order to preserve the available statistics, a
1-dimensional correction as a function of the variable under study was
used, after a careful tuning of the input kinematic distribution via
an iterative procedure.
%
%
In order to extract the inverse slope parameter $T$, the $p_{T}$
distributions were fitted with the exponential function $1/p_{T}\,
dN/dp_{T} = \exp{(-m_{T}/T)}$. The NA60 acceptance goes down to $p_{T}
= 0~\mathrm{GeV}$ and the $p_{T}$ spectra have statistics up to
$\gtrsim2.6~\mathrm{GeV}$.  The fit was repeated in the full range
$0.0~\mathrm{GeV} < p_\mathrm{T} < 2.6~\mathrm{GeV}$ and in the two
sub-ranges ($0.0~\mathrm{GeV} < p_\mathrm{T} < 1.6~\mathrm{GeV}$ and
$1.1~\mathrm{GeV} < p_\mathrm{T} < 2.6~\mathrm{GeV}$), corresponding
to the experimental windows of the NA49 and NA50 experiments,
respectively. The $\chi^2/\mathrm{ndf}$ of the fits ranges between 0.6
and 1.5.  The $T$ measured in the full range shows a rapid initial
increase with centrality, followed by a saturation. It should be
noticed that in presence of radial flow the $T$ slopes extracted from
exponential fits depend on the fit range~\cite{Heinz:2004qz}.  The
results in the sub-ranges are shown in figure~\ref{fig:tslope},
compared to the results of the other experiments. As seen in the
figure, when the fit is performed at high $p_{T}$, there is a decrease
of the average value of $T$ and a flatter trend with centrality. The
maximum difference, in the most central bin, is of the order of $\sim
15~\mathrm{MeV}$.  The discrepancy between NA50 and NA49 seems not
consistent with the $T$ slope variation observed in the NA60 data,
suggesting that some other mechanism like kaon absorption/rescattering
may be at play.  A direct comparison of the absolute values of $T$
between NA60 and the Pb-Pb experiments is not straightforward due to
possible system-size effects.

\begin{figure}[tbp]
  \centering
  \includegraphics[width=0.53\textwidth]{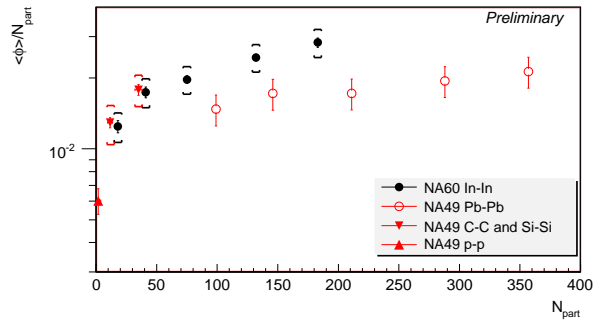}
  \caption{$\phi$ yield normalized to the number of participants in
    full phase space and corrected for the branching ratio. Compared
    to the NA49 measurement in several collision systems. The error
    bar of the Pb-Pb data includes statistical and systematic error.
    Square brackets: total error (statistical and systematic).  C, Si
    and peripheral In points displaced by $\pm 2$ participants to
    improve readability.  }
  \label{fig:yield}
\end{figure}

Two different approaches with largely independent systematics were
used to determine the average number of $\phi$ mesons produced per
interaction, $\left<\phi\right>$.  In the first, the $\phi$ cross
section integrated in centrality $\sigma_{\phi}$ has been evaluated
using the observed number of $\phi$ mesons and the number of incident
ions measured with 2 independent beam counters. $\left<\phi\right>$
was then calculated dividing $\sigma_{\phi}$ by the total inelastic
In-In cross section, estimated with the general formula
$\sigma_{inel,AB} = \pi r^2_0 \left[A^{1/3} + B^{1/3} -\beta \left(
    A^{-1/3} + B^{-1/3} \right) \right]$, where A and B are the mass
numbers of the two nuclei, $r_0 = 1.25~\mathrm{fm}$ and $\beta =
0.83$~\cite{Abreu:1999av}. In the second approach the $J/\psi$ was
used as a reference.  $\left<\phi\right>$ can be calculated from the
measured ratio $\left<\phi\right>/\left<J/\psi\right> =
(N_{\phi}/Acc_{\phi}\times\varepsilon_{rec,\phi})/
(N_{J/\psi}/Acc_{J/\psi}\times\varepsilon_{rec,J/\psi})$.  Once
corrected for anomalous and nuclear absorption, $\left<J/\psi\right>$
scales with the number of binary collisions and can be written as
$\left<J/\psi\right> = (\sigma_{NN}^{J/\psi}/\sigma_{NN})\cdot
N_{coll}$. By multiplying this value by the above ratio, one obtains
$\left<\phi\right>$.  The 2 methods agree within $\sim10\%$.
The $\phi$ enhancement is determined by normalizing the yield to the
number of participants. Figure~\ref{fig:yield} shows the ratio
$\left<\phi\right>/N_{part}$ and the comparison to the NA49
measurements in several collision
systems~\cite{Alt:2004wc,Friese:2002re,Alt:2006dk}.  The $\phi$
enhancement in In-In with respect to p-p reaches a factor of five in
the most central collisions. As seen, the observed $\phi$ enhancement
observed by NA49 is consistently smaller at any centrality. NA49
showed that the $\phi \to KK$ breaks $N_{part}$ scaling, and that
$\left<\phi\right>/N_{part}$ for central collisions and for several
collision system already saturates for Si-Si collisions. The value of
$\left<\phi\right>/N_{part}$ for central In-In collisions in the muon
channel exceeds the corresponding values for Pb-Pb collisions in the
kaon channel, suggesting that the NA60 $\mu\mu$ data do not follow the
trend measured by NA49 in the kaon channel.
A further reduction of the systematic uncertainties is currently in
progress.
%
An unambiguous comparison to the NA50 result in full phase space is
not possible, due to lack of consensus on the value of the slope
parameter in Pb-Pb collisions (see above).  An extrapolation with the
extreme hypotheses $T=220~\mathrm{MeV}$ and $T=300~\mathrm{MeV}$ leads
to much larger values with respect to In-In, even in the upper-bound
scenario $T=300~\mathrm{MeV}$.  The measurements of the CERES
experiment do not rule out differences of the order of those observed
by NA60, due to the large uncertainties on the $ee$ measurements.

In conclusion, the long-standing $\phi$ puzzle is yet to be clarified.
The on-going analysis of the $\phi \to KK$ channel in the NA60 data
may be decisive to clarify the picture.

The YerPhI group was supported by the C. Gulbenkian Foundation, Lisbon
and the Swiss Fund Kidagan.

\section*{References}

\bibliographystyle{jphysg} 
\bibliography{mfloris_qm08}

\end{document}